
\magnification 1200
\def\ct{\centerline}
\def\noi{\noindent}
\def\bi{\bigskip}

\def\ii{\'{\i}}
\baselineskip 20 pt
\vglue 1 cm
\ct {The gauge theory of the de-Sitter group and Ashtekar formulation.
\footnote{*}{\sevenrm This work was in part supported by CONACyT under
contracts 1683-E9209 and F246-E9207 and }
 \footnote {}{\sevenrm by Coordinaci\'on de Investigaci\'on
Cient\'{\i}fica de la UMSNH.}}
\bi\bi
\ct {\it J. A. Nieto \footnote{$^1$}{\sevenrm Also Instituto
de F\ii sica y Matem\'aticas de la Universidad Michoacana de San
Nicol\'as de Hidalgo,}
\footnote {} {\sevenrm ~~A.P. 749, C.P. 58000, Morelia, Michoac\'an},
O. Obreg\'on \footnote {$^2$} {\sevenrm Under agreement with
Departamento de F\ii sica, Universidad Aut\'onoma Metropolitana-Iztapalapa}
and J. Socorro$^2$}
\bi
\ct {\it Instituto de F\'{\i}sica de la Universidad de Guanajuato,}
\bi
\ct {\it Apartado Postal E-143, C. P. 37150, Le\'on, Guanajuato, }
\bi
\ct {\it M\'exico}
\vglue 1.5 cm
\ct {\bf ABSTRACT}
\bi
By adding the Pontrjagin topological invariant to the gauge theory
of the de Sitter group proposed by MacDowell and Mansouri we
obtain an action quadratic in the field-strengths, of the Chern-Simons
type, from which the Ashtekar formulation is derived.
\bi\bi\bi
\noi Submitted to Phys. Lett. A
\vfil\eject
The central idea in this work is to show that the theories of
MacDowell-Mansouri [1] and Ashtekar [2] are closely related. These
theories have been two of the most interesting approaches ever
proposed in an attempt to understand Einstein gravity as a gauge
theory. The original motivation, the construction, and even the
intention itself, makes at first sight, to look at these two theories
as unrelated. However, in this work we show that they
are, in fact, closely related. We show that by adding the Pontrjagin
topological invariant to the action proposed by MacDowell and
Mansouri, the Ashtekar theory in the way proposed by Samuel [3]
and Jacobson and Smolin [4], is derived. It is interesting that
according to this result the MacDowell and
Mansouri procedure may have been used to discover (nine years
before) the Ashtekar new variables. As a consecuence of this
relation a new insight about Ashtekar new variables is gained.
\bi
Since the theory of MacDowell and Mansouri is closely related to the
Chern-Simons [5] formalism we think that the present work is interesting
because opens the possibility to relate also Ashtekar theory to Chern-Simons
formalism. Ashtekar variables have been also obtained, in the context of
the Poincare group, by supplementing the gauge Lagrangian with the generating
function dictated by Riemann-Cartan geometry [6].
\bi
On the other hand, since
the canonical quantization of the Ashtekar theory has lead to some
authors [7] to find solutions in which the states are exponential of the
Chern-Simons action we think that it is interesting that in both
classical and quantum level the Ashtekar theory is closely related to
Chern-Simons.
\bi
The starting point in the MacDowell-Mansouri construction is to consider the
gauge potential $\omega_\mu\,^{\rm AB} (x)$ where the indice
$\mu = 0, 1, 2, 3$ corresponds to a base four dimensional space-time M,
with $\rm x^\alpha$ local coordinates on M, and the indices A,B $= 0,1,2,3,4$
are associated to the fibre anti-de-Sitter group SO(3,2) (or alternatively to
the de-Sitter group SO(4,1) in case
 that the theory is not supersymmetrized). So, the gauge potential
$\omega_\mu\,^{\rm AB}$ is a connection associated to the fibre-bundle
(M, SO(3,2)).
\bi
{}From the gauge potential $\omega_\mu\,^{\rm AB}$ we may introduce the
curvature
$$ {\cal R}^{^{\rm AB}}_{_{\mu \nu}} = {\rm \partial_\mu \omega_\nu\,^{AB}
 - \partial_\nu \omega_\mu\,^{AB} + {1 \over 2} f^{^{[AB]}}_{_{[[CD][EF]]}}
\omega_\mu\,^{CD} \omega_\nu\,^{EF}}, \eqno (1)$$
\noi
where $\partial_\mu \equiv {\partial \over \partial x^\mu} $ and
$\rm  f^{^{[AB]}}_{_{[[CD][EF]]}} $ are the structure constants of SO(3,2).
Here the notation in the indices $\rm [AB]$ means antisimmetrization. The
gauge potential itself $ \omega_{\mu}\,^{\rm AB}$ and the curvature
${\cal R}^{\rm AB}_{\mu \nu}$ should be written rigorously as
 $\omega_\mu\,^{[\rm AB]} $ and ${\cal R}^{[\rm AB]}_{[\mu \nu]}$, but it
turns out to be more convenient to keep in mind this fact instead to
write all the time the bracket [AB].
\bi
Consider, now, the generators $\rm S_{AB} = - S_{BA}$ of the
anti-de-Sitter algebra SO(3,2) satifying

$$\rm  [S_{AB}, S_{CD} ] = f^{[EF]}_{[[AB] [CD]]} S_{EF}. \eqno (2) $$

\noi
We find that for SO(3,2) the structure constants are given by

$$\rm  f^{^{[EF]}}_{_{[[AB][CD]]}} = {1 \over 2} [\eta_{AC} \delta^E_B
 \delta^F_D - \eta_{AD} \delta^E_B \delta^F_C + \eta_{BD} \delta^E_A
\delta^F_C - \eta_{BC} \delta^E_A \delta^F_D] - (E \leftrightarrow F).
\eqno (3) $$
{}From this expresion we obtain the following results:
$$\rm  f^{^{[4a]}}_{_{[4b][4c]}} = 0, \quad f^{^{[4d]}}_{_{[ab][cd]}} = 0,
\quad f^{^{[ab]}}_{_{[4c][de]}}= 0, \eqno (4)$$
$$\rm f^{^{[4a]}}_{_{[4b][cd]}} = {1 \over 2} [ \eta_{bd} \delta^a_c -
\eta_{bc}
$$\rm f^{^{[ab]}}_{_{[4c][4d]}} = {{\lambda^2} \over 2} [ - \delta^a_c
\delta^b_
$$\rm  f^{^{[ab]}}_{_{[cd][ef]}} = {1 \over 2} [ \eta_{ce} \delta_d^a
\delta^b_f
\noi
where the indices a,b run from 0 to 3. Using these results the algebra (2)
may be breaking as follows:
$$\rm [J_{ab}, J_{cd} ] = f^{[ef]}_{[ab][cd]} J_{[ef]} , \eqno (8) $$
$$\rm [P_a, J_{cd}] = \eta_{ac} P_d - \eta_{ad} P_c , \eqno (9) $$
$$\rm [P_a, P_b] = - \lambda^2 J_{ab},\eqno (10)$$
where $\rm P_a = \lambda S_{4a}$ and $\rm J_{ab} = S_{ab}$. Note that when
$\lambda \rightarrow 0$ the algebra (8)-(10) reduces to the Poincare algebra.
\bi
Let us now see what are the consecuences of substituting the results (4)-(7)
in the curvature (1). We find

$$ {\cal R}^{\rm a4}_{\mu \nu} =\rm  \partial_\mu e_{\nu}\,^a - \partial_\nu
e_{\mu}\,^a + (e_{\mu}\,^b \omega_{\nu b}\,^{a} - \omega_{\mu b}\,^a
e_\nu\,^b ) , \eqno (11) $$
and
$$ {\cal R}^{\rm ab}_{\mu \nu} =\rm R^{ab}_{\mu \nu} -  \lambda^2
(e_{\mu}\,^a e_{\nu}\,^b - e_{\nu}\,^a e_{\mu}\,^b)  , \eqno (12) $$
where $\rm e_{\mu}\,^a \equiv \omega_{\mu}\,^{4a} $ and
$$\rm R^{ab}_{\mu \nu} = \partial_\mu \omega_{\nu}\,^{ab} -
 \partial_\nu \omega_{\mu}\,^{ab} + \omega_{\mu}\,^{ca}
\omega_{\nu c}\,^b - \omega_{\nu}\,^{ca}
\omega_{\mu c}\,^b . \eqno (13) $$
It turns out that ${\cal R}^{\rm a4}_{\mu \nu} $ may be identifyed as the
torsion, while $\rm R^{ab}_{\mu \nu}$ is the usual curvature Riemann tensor
in four dimensions.
\bi
For later purpose, now, we define the self dual connections $^+\omega $ :
$$\rm  ^+\omega_{\mu}\,^{ab} = {1 \over 2} (\omega_{\mu}\,^{ab} - i
\epsilon^{ab}\,_{cd} \omega_{\mu}\,^{cd} ), \eqno (14) $$
and the antiself dual connection $^-\omega$:
$$\rm ^-\omega_\mu\,^{ab} = { 1 \over 2} (\omega_{\mu}\,^{ab} + i
\epsilon^{ab}\,_{cd} \omega_{\mu}\,^{cd} ). \eqno (15)$$
In addition, we introduce the object
$$\rm \Sigma^{ab}_{\mu \nu} = e_{\mu}\,^a e_{\nu}\,^b - e_{\nu}\,^a
e_{\mu}\,^b, \eqno (16) $$
with its  corresponding self dual
$$\rm ^+\Sigma^{ab}_{\mu \nu} = {1\over 2} (\Sigma^{ab}_{\mu \nu}-
{i\over 2} \epsilon^{ab}\,_{cd}\, \Sigma^{cd}_{\mu \nu}), \eqno (16a)$$
and antiself dual part
$$\rm ^-\Sigma^{ab}_{\mu \nu} = {1\over 2} (\Sigma^{ab}_{\mu \nu}+
{i\over 2} \epsilon^{ab}\,_{cd}\, \Sigma^{cd}_{\mu \nu}). \eqno (16b)$$
Using (14), (15) and (16) the curvature (12) becomes
$$ {\cal R}^{\rm ab}_{\mu \nu} = ^+{\cal R}^{\rm ab}_{\mu \nu} +  \,^-{\cal
R}^{
where
$$\rm  ^+{\cal R}^{ab}_{\mu \nu} = \partial_\mu ~ ^+ \omega_{\nu}\,^{ab} -
\partial_\nu ~ ^+{\omega}_{\mu}\,^{ab} + ~ ^+{\omega}_{\mu}\,^{ca} ~
^+{\omega}_{\nu c}^{~~b} - ~ ^+{\omega}_{\nu}\,^{ca} ~ ^+ \omega_{\mu c}^{~~b}
- \lambda^2 ~^+{\Sigma}^{ab}_{\mu \nu}, \eqno (18) $$
and a similar expression for $^-{\cal R}^{\rm ab}_{\mu \nu}$.
\bi
Here we are interested to take only the self dual part of
${\cal R}^{\rm ab}_{\mu \nu}$. So we have
$$ ^+{\cal R}^{\rm ab}_{\mu \nu} =\rm  ^+R^{ab}_{\mu \nu} - \lambda^2 \,
^+\Sigma^{ab}_{\mu \nu},  \eqno (19) $$
where
$$\rm ^+R^{ab}_{\mu \nu} = \partial_\mu \,  ^+\omega_\nu \,^{ab} -
\partial_\nu\,^+\omega_\mu\,^{ab} +\, ^+\omega_\mu\,^{ca}\,
 ^+\omega_{\nu c}\,
^b -\, ^+\omega_\nu\,^{ca} \, ^+\omega_{\mu c}\,^b. \eqno (20) $$
\bi
Consider now the action
$$\rm  S = \int d^4 x\, \epsilon^{\mu \nu \alpha \beta}\,
 \epsilon_{abcd} \,
 ^+{\cal R}^{ab}_{\mu \nu}\,  ^+{\cal R}^{cd}_{\alpha \beta}.
 \eqno (21) $$
\noi
Introducing (19) into (21) we find
$$ \eqalign {\rm S & = \int{\rm d^4 x}\, \epsilon^{\mu \nu \alpha \beta}\,
\epsilon_{\rm abcd}\,  ^+{\rm R^{ab}_{\mu \nu}}\,
^+{\rm R^{cd}_{\alpha \beta}} \cr
& - 2 \lambda^2 \int {\rm d^4 x}\, \epsilon^{\mu \nu \alpha \beta}\,
 \epsilon_{\rm abcd} \,^+\Sigma^{\rm ab}_{\mu \nu}\,
 ^+{\rm R^{cd}_{\alpha \beta}} \cr
& + 2 \lambda^4 \int {\rm d^4 x}\, \epsilon^{\mu \nu \alpha \beta}\,
\epsilon_{\rm abcd}\, {\rm e_\mu\,^a}\, {\rm e_\nu\,^b}\,
{\rm e_\alpha\,^c}\, {\rm e_\beta\,^d, }} \eqno (22) $$
where we used (16) in order to obtain the last term in (22).
We recognize the second term as the Ashtekar action [3,4], while the last
expression is the usual cosmological term. So, the second and the third term
corresponds to Ashtekar theory.
\bi
Let us now clarify the meaning of the first term in (22). Using (14) we see
that $\rm ^+R^{ab}_{\mu \nu}$ given in (20) may be written as
$$\rm ^+R^{ab}_{\mu\nu} = {1 \over 2} (R^{ab}_{\mu \nu} -
{i \over 2} \epsilon^{ab}\,_{cd} R^{cd}_{\mu \nu} ). \eqno (23) $$
So, the first term in (22) becomes
$$ \eqalign { & {1 \over 4} \int {\rm d^4 x}\,
 \epsilon^{\mu \nu \alpha \beta}\, \epsilon_{\rm abcd}
 \bigg ( {\rm R^{ab}_{\mu \nu}} -
 {i\over 2} \epsilon^{\rm ab}\,_{\rm ef} {\rm R^{ef}_{\mu \nu}} \bigg )
\cdot \bigg ( {\rm R^{cd}_{\alpha \beta}} -
{i\over 2} \epsilon^{\rm cd}\,_{\rm gh}
{\rm R^{gh}_{\alpha \beta}} \bigg ) \cr
& = {1 \over 4} \int {\rm d^4 x}\, \epsilon^{\mu \nu \alpha \beta}\,
\epsilon_{\rm abcd}\, {\rm R^{ab}_{\mu \nu}}\,
{\rm R^{cd}_{\alpha \beta}} -  { 1 \over 16 } \int {\rm d^4 x}\,
\epsilon^{\mu \nu \alpha \beta}\, \epsilon_{\rm abcd}\,
 \epsilon^{\rm ab}\,_{\rm ef}\, \epsilon^{\rm cd}\,_{\rm gh}
{\rm R^{ef}_{\mu \nu}}\, {\rm R^{gh}_{\alpha \beta}} \cr
&- {i\over 4} \int {\rm d^4 x}\, \epsilon^{\mu \nu \alpha \beta}\,
\epsilon_{\rm abcd}\,
{\rm R^{ab}_{\mu \nu}}\, \epsilon^{\rm cd}\,_{\rm gh}
{\rm R^{gh}_{\alpha \beta}}. \cr } \eqno (24) $$
Considering that
$$\rm \epsilon_{abcd}\, \epsilon^{ab}\,_{ef} = - 2
 (\eta_{ce} \eta_{df}- \eta_{cf}  \eta_{de}), \eqno (25) $$
\bi
\noi
the second term in (24) yields
$$ \eqalign { & + {1 \over 8} \int {\rm d^4 x}\,
\epsilon^{\mu \nu \alpha \beta}\, (\eta_{\rm ce}
\eta_{\rm df} - \eta_{\rm cf} \eta_{\rm de})
\epsilon^{\rm cd}\,_{\rm gh} {\rm R^{ef}_{\mu \nu}}
{\rm R^{gh}_{\alpha \beta}} \cr
& = {1 \over 4} \int {\rm d^4 x}\, \epsilon^{\mu \nu \alpha \beta}\,
\epsilon_{\rm efgh} {\rm R^{ef}_{\mu \nu}} {\rm R^{gh}_{\alpha \beta}}.
\cr }\eqno (26)$$
\bi
\noi
Therefore the first two terms in (24) reduce to

$$ {1 \over 2} \int {\rm d^4 x}\, \epsilon^{\mu \nu \alpha \beta}\,
\epsilon_{\rm abcd}\, {\rm R^{ab}_{\mu \nu}}\,
{\rm R^{cd}_{\alpha \beta}}. \eqno (27) $$

Now, the last term in (24) leads to

$$ \eqalign { & - {i \over 4} \int {\rm d^4x}\,
\epsilon^{\mu \nu \alpha \beta}\, \epsilon_{\rm abcd}\,
\epsilon^{\rm cd}\,_ {\rm gh} {\rm R^{ab}_{\mu \nu}}
{\rm R^{gh}_{\alpha \beta}} \cr
& = + {i \over 2} \int {\rm d^4 x}\, \epsilon^{\mu \nu \alpha \beta}\,
(\eta_{\rm ag} \eta_{\rm bh} - \eta_{\rm ah} \eta_{\rm bg} )
{\rm R^{ab}_{\mu \nu}}\, {\rm R^{gh}_{\alpha \beta}} \cr
& = i  \int {\rm d^4x}\, \epsilon^{\mu \nu \alpha \beta}\,
{\rm R^{ab}_{\mu \nu}}\, {\rm R^{gh}_{\alpha \beta}}\,
\eta_{\rm ag} \eta_{\rm bh}. \cr } $$
\noi
Sumarizing the first term in (22) leads to the result
$$ \eqalign {  & \int {\rm d^4 x}\,
\epsilon^{\mu \nu \alpha \beta}\,
\epsilon_{\rm abcd}\, ^+{\rm R^{ab}_{\mu \nu}}\,
 ^+{\rm R^{cd}_{\alpha \beta}} \cr
& = {1 \over 2} \int {\rm d^4 x}\,
\epsilon^{\mu \nu \alpha \beta}\, \epsilon_{\rm abcd}
{\rm R^{ab}_{\mu \nu}} {\rm R^{cd}_{\alpha \beta}} \cr
&+ i \int {\rm d^4 x}\, \epsilon^{\mu \nu \alpha \beta}\,
\eta_{\rm ac} \eta_{\rm bd}\, {\rm R^{ab}_{\mu \nu}}\,
{\rm R^{cd}_{\alpha \beta}}.  \cr } \eqno (28) $$
We recognize the first term as the Euler topological term and the second is
the Pontrjagin topological term. So, the first term in (22) is a topological
term.
\bi
It is known [3,4] that the second term in (22) gives half the
Einstein-Hilbert action. Therefore, considering the second and
the third terms in (22) and the
first term in (28) we get the approach proposed by MacDowell and Mansouri
[1]. The new feature comes from the second term in (28). In other words, by
adding
the second term in (28) to the MacDowell and Mansouri action and by adding to
the Einstein-Hilbert action a term which is identically zero we may be able
to get an action which reduces to both the gravity gauge theory of MacDowell
and Mansouri and the Ashtekar formalism with cosmological constant and two
extra topological terms.
\bi
Let us conclude this paper with some few remarks. In Ashtekar formalism the
connection $\rm  ^+\omega_\mu\,^{ab}$ usually called $\rm A_\mu\,^{ab}$ plays
a central role. In the literature, the field strenght associated to
$\rm A_\mu\,^{ab}$ is called $\rm F^{ab}_{\mu \nu}$ which is defined as

$$\rm  F^{ab}_{\mu \nu} = \partial_\mu A_\nu\,^{ab} -
\partial_\nu A_\mu\,^{ab} + \ A_\mu\,^{ca}\,A_{\nu c}\,^b - A_\nu\,
^{ca}\,A_{\mu c}\,^b. \eqno (29) $$
In our approach $\rm F^{ab}_{\mu \nu} = ~ ^+R^{ab}_{\mu \nu}$. Just because
$\rm F^{ab}_{\mu \nu}$ looks like a Yang-Mills field strenght Ashtekar theory
is assumed ordinarily as a Yang-Mills version of gravity. However, the action
in Ashtekar formulation is
$$\rm  \int  d^4 x\, \epsilon^{\mu \nu \alpha \beta}\, \epsilon_{ abcd}\,
 e_\mu\,^a\,  e_\nu\,^b F^{cd}_{\alpha \beta}, \eqno (30) $$
which has not resemblance with a Yang-Mills action. So, from this point of
view the Ashtekar formulation does not look a like a gauge field theory of
gravity. However, in this paper we have shown that by adding three terms to
the action (30) (cosmological, Euler and Pontrjagin terms) we may arrive
to the action (21) which has Yang-Mills prototype form. Therefore, the action
(21) written as
$$\rm  S= \int d^4 x \epsilon^{\mu \nu \alpha \beta} \epsilon_{abcd}
{\cal F}^{ab}_{\mu \nu} {\cal F}^{cd}_{\alpha \beta}, \eqno (31) $$
with ${\cal F}^{\rm ab}_{\mu \nu} = {\rm F^{ab}_{\mu \nu}} - \lambda^2 ~^+
\Sigma^{\rm ab}_{\mu \nu}$ provide in our opinion the correct argument to say
that
gravity can be treated as a gauge theory. From this point of view we
think that in part the Ashtekar formalism has been clarified.
\bi
Finally, it is known that actions of the form (31) are closely related to
Chern-Simons. In fact if instead of the Levi-Civita  $\epsilon_{\rm abcd}$
tensor we use the Killing metric associated to the de-Sitter Group
SO(3,2), expression (30) becomes the Chern-Simons action. On the other hand,
it is also known [7] that canonical quantization of the Ashtekar theory leads
to constraints whose solution is of the  form
$\psi \sim \exp({{\rm const.}\over \Lambda}S_{cs})$. So, according to our
result Ashtekar formalism seems to be closely
related to Chern-Simons no only classically but also producing the physical
states of  quantum general relativity. The quantum canonical formulation of
the action (21) as well as the analysis of its consequences in the
corresponding wave function of gravity will be reported elsewhere.
\vfil\eject
\ct {\bf REFERENCES}
\bi
\item {[1]} S. W. MacDowell and F. Mansouri, Phys. Rev. Lett. 38 (1977) 739;
F. Mansouri, Phys. Rev. D16 (1977) 2456; P. Van Nieuwenhuizen, Phys. Rep. 68
(1981) 189-398; P. G. O. Freud, ``Introduction to Supersymmetry" Cambridge
University Press 1986.

\item {[2]} A. Ashtekar, Phys. Rev. Lett. 57 (1986) 2244; Phys. Rev. D36
(1987) 1587. See A. Ashtekar ``New perspectives in canonical Gravity" 1988.
Monographs and textbooks in physical science, Napoli, Bibliopolis, and
references there in.

\item {[3]} J. Samuel, Pramana J. Phys. 28 (1987) L429.

\item {[4]} T. Jacobson and L. Smolin, Class. Quant. Grav. 5 (1988) 583.

\item {[5]} S. S. Chern and J. Simons, Ann. Math. 99 (1974) 48.

\item {[6]} E. W. Mielke, Phys. Lett. A149 (1990) 345; Phys. Rev. D42 (1990)
3388; P. K. Townsend, Phys. Rev. D15 (1977) 2795.

\item {[7]} See B. Br\"ugmann, R. Gambini and J. Pullin, Gen. Rel. and Grav.
25, (1993) 1  and references there in.

\end